\documentclass[pra,aps,twocolumn,showpacs,10pt]{revtex4-1}
\usepackage{amsmath,amssymb,graphicx,amsthm,dsfont,bm,color,ulem}

\begin{document}

\title{Effects of orbital angular momentum on the pulse shape at the most intense ring of ultrafast vortices}

\author{Miguel A. Porras}

\affiliation{Grupo de Sistemas Complejos, ETSIME, Universidad Polit\'ecnica de Madrid, Rios Rosas 21, 28003 Madrid, Spain}

\begin{abstract}
It has been recently shown that the orbital angular momentum (OAM) and temporal degrees of freedom in ultrafast (few-cycle) vortices are coupled. This coupling manifests itself with different effects in different parts of the vortex, as has been shown for the ring surrounding the vortex where the pulse energy is maximum, and also in the immediate vicinity of the vortex center. In may applications, however, the ring of maximum energy is not of primary interest, but that where the pulse peak intensity is maximum, which is particularly true in nonlinear optics applications such as the experiments with ultrafast vortices exciting high harmonics and attosecond pulses carrying also OAM. This article describes the effects of the OAM-temporal coupling at the ring of maximum pulse peak intensity, which does not always coincide with the ring of maximum pulse energy. We find that there exists an upper bound to the magnitude of the topological charge that an ultrafast vortex with prescribed pulse shape at its most intense ring can carry, and vice versa, a lower bound to the duration of the pulse at the most intense ring for a given magnitude of the topological charge. These bounds imply that with a given laser source spectrum, the duration of the synthesized ultrafast vortex increases with the magnitude of the topological charge. Explicit analytical expressions of ultrafast vortices containing these OAM-temporal couplings are given that can be of interest in a variety of applications, particularly in the study of their propagation and interaction with matter.
\end{abstract}

\maketitle

\section{Introduction}

Although continuous light can transport vortices with arbitrarily high magnitude of the topological charge $l$, and therefore arbitrarily high orbital angular momentum (OAM) (see, e.g., \cite{ALLEN,YAOPADGETT}), pulsed light cannot \cite{PORRAS_PRL,CONTI_PRL}. This is a consequence of the coupling between the OAM and temporal degrees of freedom in pulsed vortices, as described also in Refs. \cite{PORRAS_OL,CONTI_PRA}.

Theoretically, this coupling affects pulsed vortices of any duration and topological charge, but it is only with further advances of the current technology in the generation of shorter and shorter vortices \cite{BEZUHANOV,MARIYENKO,BEZUHANOV2,ZEYLIKOVICH,TOKIZANE,SHVEDOV,RICHTER,YAMANE,MIRANDA} of high quality (without topological charge and angular dipersions), at high powers, and approaching the single-cycle regime, and the use of these ultrafast vortices in strong-field light-matter interactions for the generation of high harmonics and attosecond pulses with higher and higher OAM \cite{HERNANDEZ,GARIEPY,REGO,TURPIN,REGO2}, that the effects of this coupling will be observable. The OAM-temporal coupling in these ultrafast vortices with high OAM will then have an impact in ultrafast vortex applications such as classical and quantum information and entanglement \cite{GIBSON,HUANG,MAIR,MOLINA}, transfer of OAM to matter \cite{HE}, material processing \cite{OMATSU} or nanosurgery \cite{JEFFRIES}. Understanding the effects of OAM-temporal coupling is also fundamental when using the chirality of these vortices to probe chiral properties of matter at ultrashort time scales \cite{AYUSO,SILVA}.

Spatiotemporal couplings in ultrashort, pulsed light beams have been known for a long time (see, e. g., \cite{AKTURK}), particularly in ultrashort, fundamental Gaussian beams in their various forms \cite{SHEPPARD,KAPLAN,KAPLAN,PORRAS_PRE,PORRAS_JOSAB,PORRAS_PRE3}. OAM-temporal coupling is a particular case of spatiotemporal coupling, say an azimuthal-temporal coupling, which seems to be rather more pronounced than the radial-temporal coupling in ultrashort Gaussian beams. In \cite{PORRAS_PRL,CONTI_PRL,PORRAS_OL,CONTI_PRA} the effects of OAM-temporal coupling have been described for the first time in ultrafast vortices shaped as pulsed Laguerre-Gauss (LG) beams, commonly used in the experiments, and also shaped as pulsed Bessel beams, or diffraction-free X-waves. These effects are reviewed in \cite{PORRAS_PRA2}, where it turns out that the OAM-temporal coupling manifests itself with different effects in different parts of the vortex. With a given broadband laser source, the carrier frequency experiences an important blue shift with respect to the laser mean frequency, $\bar\omega$ (defined in the standard way  \cite{BRABEC} as the centroid of the spectral density) in the immediate vicinity of the vortex center, accompanied by an increment the number of oscillations, and these effects are more pronounced as the magnitude of the topological charge is higher \cite{CONTI_PRL,PORRAS_PRA2}. On the contrary, in the ring surrounding the vortex where the pulse energy is maximum, or ``bright" ring for a time-integrating detector, the mean carrier frequency is approximately the same as that of the laser source regardless of the topological charge, and the number of oscillations and duration of the pulse, $\Delta t$ (defined by means of the central second-order moment of the pulse intensity), are larger compared to that obtainable with the laser source spectrum, and they increase with the magnitude of the topological charge, in such a way that the duration is always above the minimum possible duration, $\sqrt{|l|}/\bar\omega$, for ultrafast vortices of charge $l$ \cite{PORRAS_PRL,PORRAS_OL,PORRAS_PRA2}.

We note in this paper that the ring of the ultrafast vortex where the pulse energy is maximum is not always of primary interest. High pulse energy may be the result of a long duration with relatively low peak intensity, while pulses as short and as intense as possible (resulting probably in lower energy) are typical demands in ultrafast nonlinear optics experiments, particularly in strong-field laser-matter interactions \cite{HERNANDEZ,GARIEPY,REGO,TURPIN,REGO2}. In these situations, minimum duration with a given laser source spectrum is obtained with uniform spectral phases, i. e., with transform-limited or unchirped pulses. This is why we limit our considerations to transform-limited pulses, and describe the effects of OAM-temporal coupling at the ring of the ultrafast vortex where the pulse peak intensity is maximum, which generally does not coincide with the ring of maximum pulse energy, as demonstrated below.

It turns out that the effects of OAM on the transform-limited pulse with maximum peak intensity are directly related to the properties of its symmetric temporal lobe around the pulse peak, which is, on the other hand, the only portion of the pulse of interest in nonlinear optics applications. Specifically, the involved magnitudes are the instantaneous frequency (derivative of the phase of the pulse with respect to time) at the instant of pulse peak, or ``central" frequency, $\omega_c$, and the duration of the central lobe around the pulse peak, defined through the lobe concavity, and that we call ``central" duration, $\Delta t_c$. As seen below, these two magnitudes are trivially determined from a measurement of the pulse spectrum.

We then find qualitatively similar OAM-coupling effects in the most intense ring to those occurring in the most energetic ring. The central frequency of the pulse at the most intense ring is the same as the central frequency of the ultrafast laser source, and therefore independent of the imprinted topological charge. The duration of the central lobe is larger than that obtainable from the laser source without OAM, and increases with the magnitude of the topological charge. This effect is a result of the existence of a lower bound, $\sqrt{|l|}/\bar\omega_c$, to the duration of the central temporal lobe that is satisfied by the pulse at the most intense ring by all existing ultrafast vortices. If, as in many situations, the ring of maximum pulse energy coincides with the ring of maximum pulse peak intensity, then the pulse temporal shape at this ring is always such that the two inequalities, $\Delta t>\sqrt{|l|}/\bar\omega$ for the pulse duration as a whole, and $\Delta t_c >\sqrt{|l|}/\bar\omega_c$ for the duration of the central temporal lobe, are satisfied.

\section{Ultrafast vortices and previous results}

We express an ultrafast vortex of topological charge $l$, or $l$ units of orbital angular momemtum, as the superposition
\begin{equation}\label{AN}
E(r,\phi,z,t')= \frac{1}{\pi}\int_0^{\infty} \hat E_\omega(r,\phi,z)e^{-i\omega t'} d\omega\, ,
\end{equation}
of monochromatic LG beams
\begin{equation}\label{LG}
\hat E_\omega(r,\phi,z) = \hat a_\omega D(z,\phi)\left[\frac{\sqrt{2}r}{s_\omega(z)}\right]^{|l|}e^{\frac{i\omega r^2}{2cq(z)}} \, ,
\end{equation}
of the same topological  charge $l$, zero radial order, and different angular frequencies $\omega$ with weights $\hat a_\omega$. In the above expressions $(r,\phi,z)$ are cylindrical coordinates, $c$ is the velocity of light in vacuum, and $t'=t-z/c$ is the local time for a plane pulse. The factor $D(z,\phi)=e^{-i(|l|+1)\tan^{-1}(z/z_R)}e^{-il\phi}/[1+\left(z/z_R\right)^2]^{1/2}$, where $z_R$ is the Rayleigh distance, accounts for the azimuthal phase variation and the $z$-dependent amplitude attenuation and Gouy's phase shift associated with diffraction. The complex beam parameter is $q_\omega(z)=z-iz_R$, and is usually written in the form
\begin{equation}\label{Q}
\frac{1}{q(z)} = \frac{1}{R(z)} + i \frac{2c}{\omega s^2_\omega (z)}\, ,
\end{equation}
where
\begin{equation}\label{S}
s_\omega(z)=s_\omega\sqrt{1+\left(\frac{z}{z_R}\right)^2}\,,\quad s_\omega=\sqrt{\frac{2z_R c}{\omega}}\,,
\end{equation}
are the width and waist width, respectively, of the fundamental ($l=0$) Gaussian beam, and $1/R(z)= z/(z^2+ z_R^2)$ is the curvature of the wave fronts of Gaussian and LG beams. Being limited to positive frequencies, the optical field $E$ in Eq. (\ref{AN}) is the analytical signal complex representation of the real field $\mbox{Re}\, E$ \cite{BORN}. As in preceding works \cite{PORRAS_PRL,PORRAS_OL,PORRAS_PRA2}, the Rayleigh distance $z_R$ is assumed to be independent of frequency, i. e., we adopt the so-called isodiffracting model \cite{PORRAS_PRE,PORRAS_JOSAB,PORRAS_PRE2,PORRAS_PRE2,FENG}, because it is the only situation in which the pulse temporal shape does not experience changes with propagation distance $z$ (except for the global complex amplitude $D$) with arbitrarily high $l$, as shown recently \cite{PORRAS_PRL}, as desired for applications. This choice is also a way to isolate the pure effects of OAM on temporal shape, i. e., the OAM-temporal couplings, and thus distinguish them from propagation effects on pulse shape in models other than the isodiffracting model, whose study is deferred to furture work.

For a simpler analysis we introduce the normalized radial coordinate $\rho = r/\sqrt{2z_R c [1+(z/z_R)^2]}$. A constant value of $\rho$ represents a revolution hyperboloid, or caustic surface, expanding as the monochromatic LG beams do. Equation (\ref{AN}) with Eq. (\ref{LG}) now reads
\begin{equation}\label{AN2}
E(\rho,\tau) = \frac{1}{\pi} \int_0^\infty \hat E_\omega(\rho) e^{-i\omega \tau}d\omega\,,
\end{equation}
where
\begin{equation}\label{LG2}
\hat E_\omega (\rho) = \hat a_\omega  D \left(\sqrt{2}\rho\right)^{|l|} \omega^{|l|/2} e^{-\rho^2\omega}
\end{equation}
and where $\tau = t' - r^2/2cR(z)$ is the local time for the ultrafast vortex. The pulse front $\tau=0$ determines the time of arrival of the pulse at each position of space, and is the same family of spherical surfaces (in the paraxial approximation) as the phase fronts of the superposed isodiffracting LG beams. According to Eqs. (\ref{AN2}) and (\ref{LG2}) the pulse temporal shape depends on the particular caustic surface $\rho$, but not on propagation distance.

In \cite{PORRAS_PRL}, the caustic surface $\rho_F$ where the time-integrated intensity, energy density, or fluence, $F(\rho) =\int_{-\infty}^\infty (\mbox{Re} E)^2 d\tau = (1/2)\int_{-\infty}^\infty |E|^2 d\tau = (1/\pi)\int_0^\infty |\hat E_\omega|^2 d\omega$ is maximum, i. e., the bright ring as recorded by a time-integrating detector, is considered the most relevant caustic surface, and the effects of OAM on pulse temporal shape at that caustic surface are described. The caustic surface of maximum fluence was shown to be determined by $\rho_F^2=|l|/2\omega(\rho_F)$, where
\begin{equation}\label{MEAN}
\bar\omega(\rho) = \frac{\int_0^\infty |\hat E_\omega(\rho)|^2 \omega d\omega}{\int_0^\infty |\hat E_\omega(\rho)|^2 d\omega}
\end{equation}
defines, as in \cite{BRABEC}, the mean frequency of the pulse at $\rho$. At this caustic surface the pulse temporal bandwidth and the topological charge were shown to be restricted by inequality $\Delta\omega(\rho_F)/\bar\omega(\rho_F)< 2/\sqrt{|l|}$, where
\begin{equation}\label{BANDWIDTH}
\Delta\omega^2(\rho) = 4 \frac{\int_0^\infty |\hat E_\omega(\rho)|^2 [\omega-\bar\omega(\rho)]^2 d\omega}{\int_0^\infty |\hat E_\omega(\rho)|^2 d\omega}
\end{equation}
is the so-called Gaussian-equivalent half-bandwidth at a caustic $\rho$ (yielding the $1/e^2$-decay half-bandwidth for a Gaussian-like spectral density $|\hat E_\omega(\rho)|^2$). Defining correspondingly the Gaussian-equivalent half-duration
\begin{equation}\label{DURATION}
\Delta t^2(\rho) = 4 \frac{\int_{-\infty}^\infty |E_(\rho,\tau)|^2 [\tau -\bar\tau(\rho)]^2 d\tau}{\int_{-\infty}^\infty |E(\rho,\tau)|^2 d\tau}\,,
\end{equation}
where
\begin{equation}
\bar\tau(\rho) = \frac{\int_{-\infty}^\infty |E_(\rho,\tau)|^2 \tau d\tau}{\int_{-\infty}^\infty |E_\omega(\rho,\tau)|^2 d\tau}\,,
\end{equation}
and on account that $\Delta t(\rho_F)\Delta\omega(\rho_F)\ge 2$ for any pulse shape, it follows that the pulse duration at the bright ring of any ultrafast vortex always satisfies $\Delta t(\rho_F)\ge \sqrt{|l|}/\bar\omega(\rho_F)$, which settles a lower bound to the duration of an ultrafast vortex with $l$ units of OAM. It has been shown later \cite{PORRAS_OL,PORRAS_PRA2} that the mean frequency at $\rho_F$ is approximately equal to the mean frequency of $\hat a_\omega$, which is more directly related to the ultrashort laser source spectrum, i. e., $\bar\omega(\rho_F)\approx \bar\omega$, where
\begin{equation}\label{MEANa}
\bar\omega = \frac{\int_0^\infty |\hat a_\omega|^2 \omega d\omega}{\int_0^\infty |\hat a_\omega|^2 d\omega}\,.
\end{equation}
With $\bar\omega(\rho_F)\approx \bar\omega$ independent of the topological charge $l$, the approximate lower bound $\Delta t(\rho_F)\gtrsim  \sqrt{|l|}/\bar\omega$ with the right hand side growing monotonously with $|l|$, the duration at the bright ring of the synthesized ultrafast vortex must necessarily increase with $|l|$ \cite{PORRAS_OL}, if the laser source spectrum $\hat a_\omega$ is fixed. These effects of the coupling between the OAM and the temporal degrees of freedom at the most energetic ring have been recently reviewed in \cite{PORRAS_PRA2}, where they are also demonstrated to be substantially the same for other types of ultrafast vortices, as OAM-carrying nondiffracting X-waves \cite{PORRAS_PRA2,CONTI_PRL,CONTI_PRA}.

\section{OAM-temporal couplings at the most intense ring of transform-limited ultrafast vortices}

\begin{figure}[tbp]
\includegraphics*[width=6.5cm]{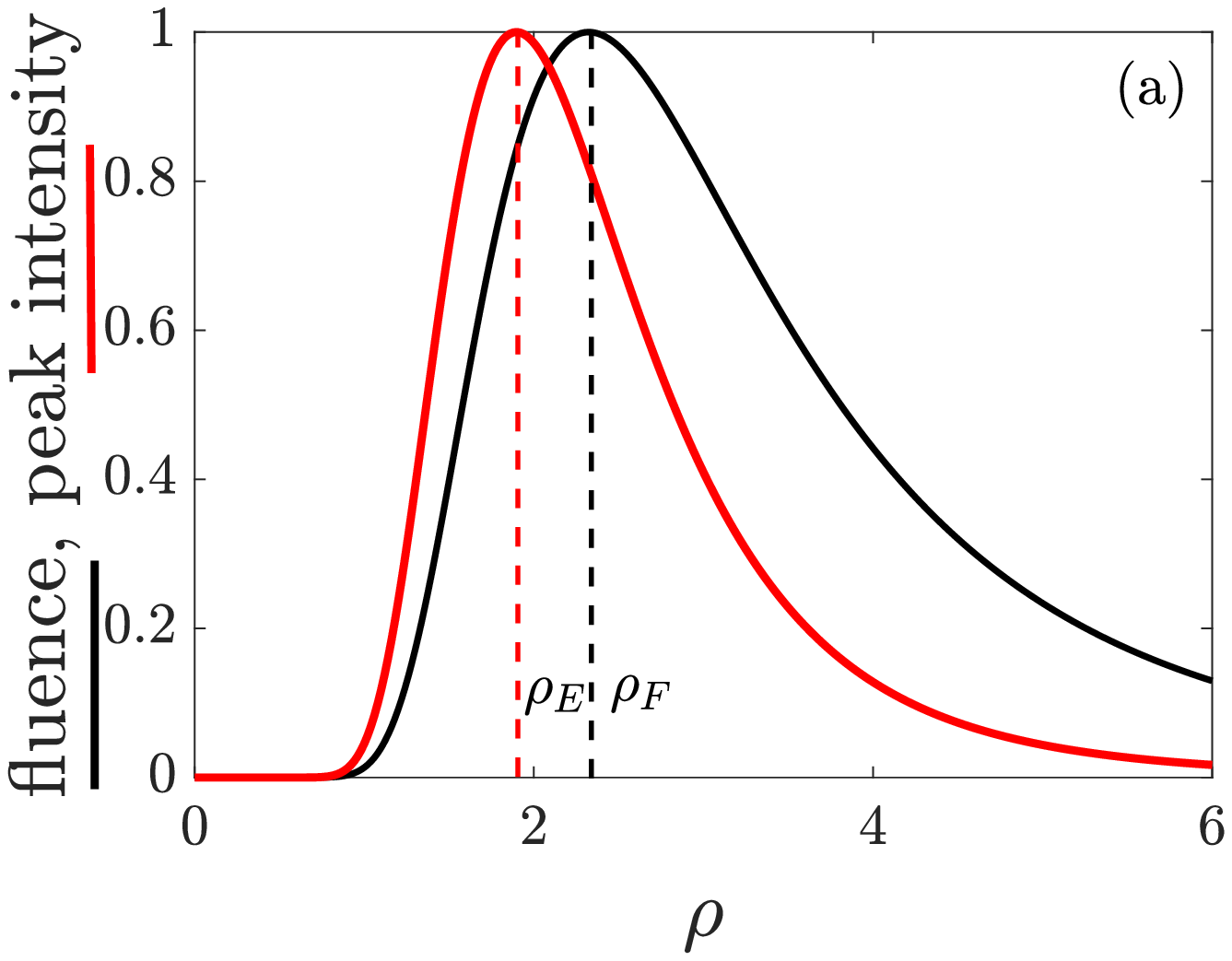}
\includegraphics*[width=6.5cm]{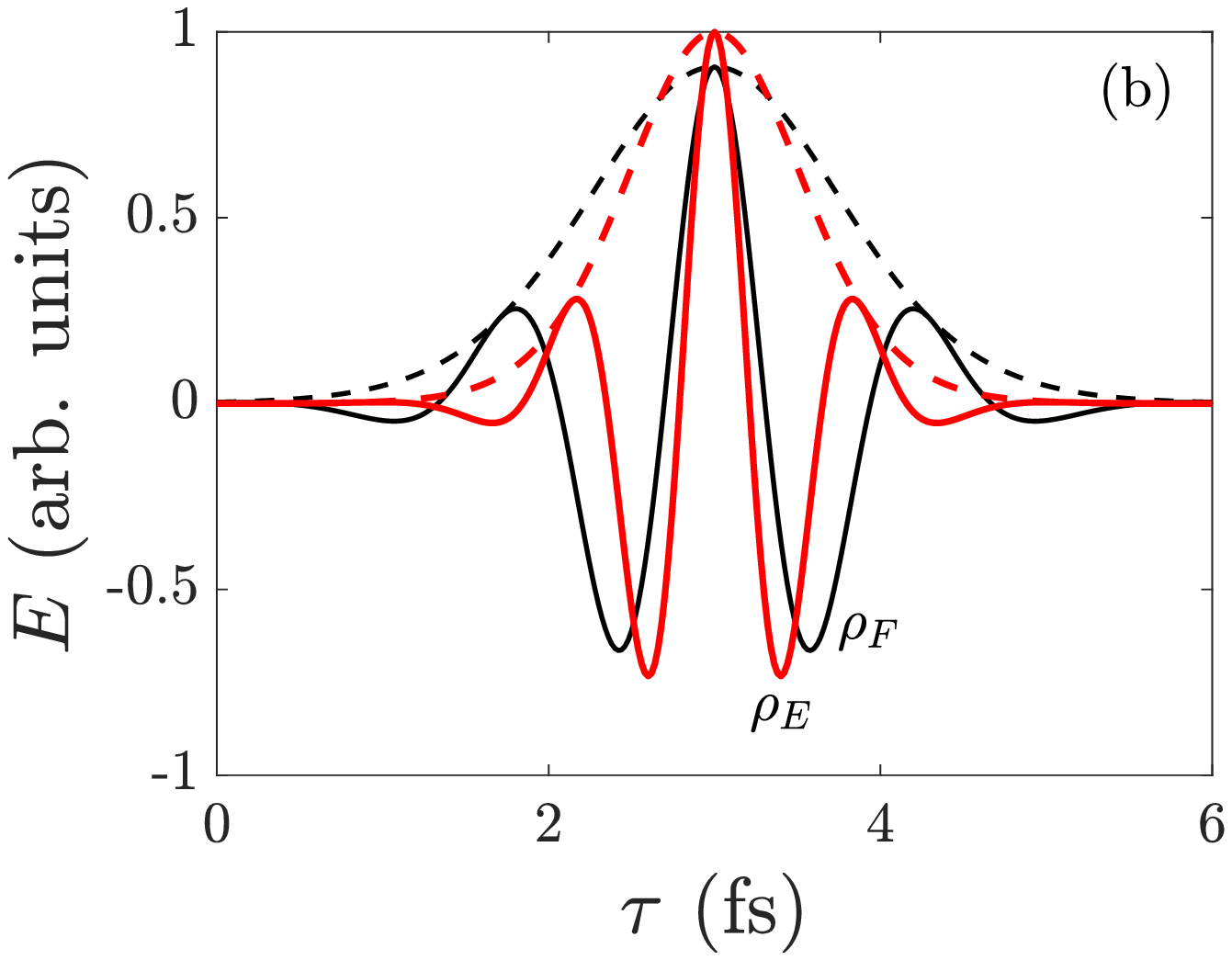}
\caption{(a) Radial profile of fluence, $F$, and peak intensity, $|E(\rho,0)|^2$, of the ultrafast vortex with topological charge $l=27$ and spectrum $\hat a_\omega = \omega^{0.5}e^{-\omega/2.5}$ of mean frequency $\bar\omega=2.5$ fs$^{-1}$, normalized to their respective maxima at the respective caustic surfaces $\rho_F=2.32$ fs$^{1/2}$ and $\rho_E= 1.90$ fs$^{1/2}$. (b) Respective pulse shapes (real field and amplitudes) at these two caustic surfaces, oscillating at different carrier frequencies, $\bar\omega (\rho_F)=2.5$ fs$^{-1}\simeq \bar \omega$ and $\bar\omega(\rho_E) =3.616$ fs$^{-1}$, and with different durations. For a better comparison, the four curves are normalized to the peak amplitude of the pulse at $\rho_E$.}
\label{Fig1}
\end{figure}

We point out here that the pulse of maximum energy is not always the ring of primary interest. In many applications, particularly in nonlinear optics applications such as the high harmonics and attosecond pulses excited by ultrafast vortices, the outcome of experiments is primarily determined by the intensity, and possibly the carrier-envelope phase. Figures 1(a) and (b) illustrate that the caustic surface where is the most energetic and the caustic surface where the pulse has maximum peak intensity are generally different, and that the pulse shapes at these two caustic surfaces, including their mean frequencies and durations, are generally different.

When the primary interest is maximum peak intensity, transform-limited pulses are desired because they have minimum duration with the available bandwidth, and then maximum intensity. The frequency spectrum $\hat E_\omega =\hat a_\omega D(\sqrt{2}\rho)^{|l|} \omega^{|l|/2}e^{-\rho^2\omega}$ at any particular point of the ultrafast vortex has uniform spectral phases, and hence the pulse $E(\rho,\tau)$ is transform-limited if the laser source spectrum $\hat a_\omega$ has uniform phases, e. g., $\hat a_\omega =\hat b_\omega e^{-i\Phi}$, and then
\begin{equation}\label{LG3}
\hat E_\omega =\hat b_\omega e^{-i\Phi} D(\sqrt{2}\rho)^{|l|} \omega^{|l|/2}e^{-\rho^2\omega}\,,
\end{equation}
where $\hat b_\omega$ can be taken as real and non-negative, and $\Phi$ determines the carrier-envelope phase. Under these conditions the pulse $E(\rho,\tau)$ consists on oscillations under an amplitude $|E(\rho,\tau)|$ that contains a symmetric central lobe about its peak value taking place at $\tau=0$, and of value
\begin{equation}\label{PEAK}
|E(\rho,\tau=0)| = |D| \left(\sqrt{2}\rho\right)^{|l|} \frac{1}{\pi} \int_0^\infty \hat b_\omega \omega^{|l|/2} e^{-\rho^2 \omega} d\omega \, .
\end{equation}
With $l\neq 0$, the peak amplitude in Eq. (\ref{PEAK}) vanishes at $\rho=0$ and for $\rho\rightarrow\infty$, so that there exists a particular caustic surface, say $\rho_E$, where the maximum pulse amplitude is also maximum transversally. The purpose of the following analysis is to describe the effects of the effects of OAM-temporal coupling at the caustic surface $\rho_E$ of ultrafast vortices.

\subsection{Pulse characterization}

\begin{figure}[b]
\includegraphics*[width=6.5cm]{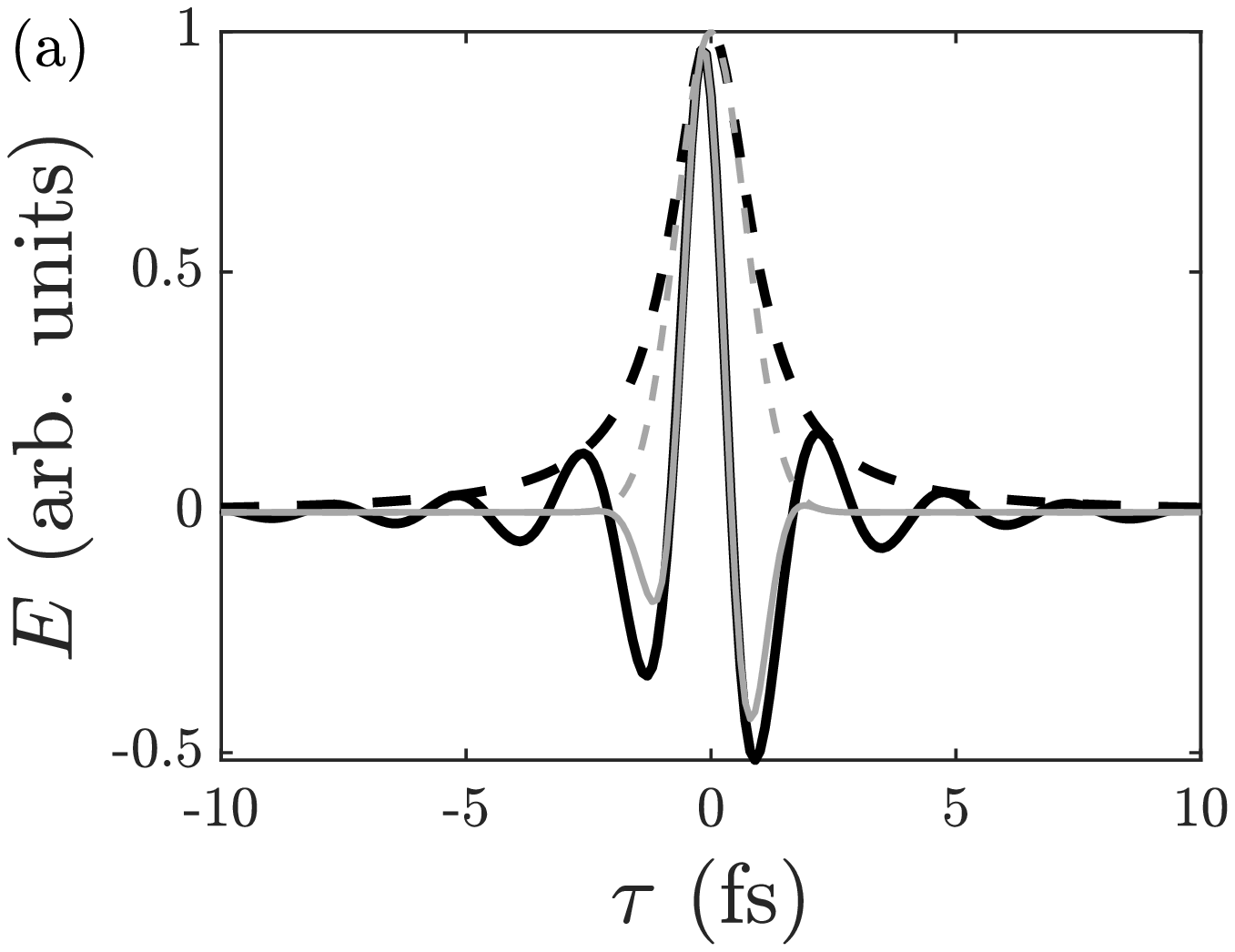}
\includegraphics*[width=6.5cm]{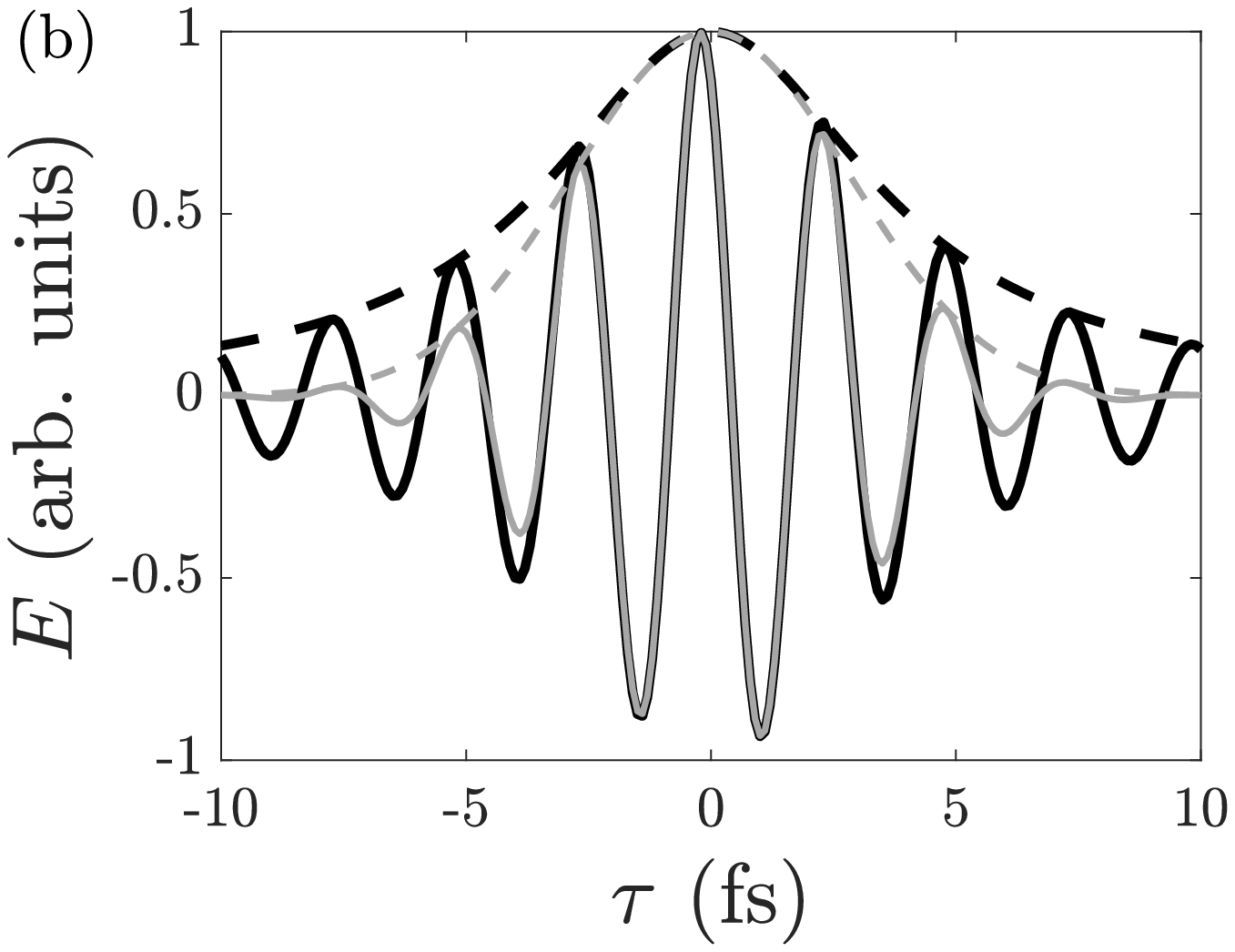}
\caption{The pulse $E=[1/(1+\tau^2/T^2)]e^{-i(\omega_0 \tau +\Phi)}$ with $\omega_0=2.5$ rad fs$^{-1}$, $\Phi=\pi/6$, and (a) $T=1$ fs and (b) $T=4$ fs (solid black curves), their amplitudes $|E|$ (dashed black curves), and their approximations $E_c=e^{-\tau^2/\Delta t_c^2}e^{-i(\bar\omega_c\tau+\Phi)}$ (gray solid curves) and $|E_c|$ (dashed gray curves) based on the central frequency $\omega_c=2.5$ fs$^{-1}$ and the central width $\Delta t_c=T$.}
\label{Fig2}
\end{figure}

As a preliminary, we introduce a characterization of transform-limited pulses by means of two measurable parameters that will be seen to be  directly involved in the OAM-temporal couplings, and appears to be relevant to nonlinear applications with these pulses, since it describes quite accurately the shape of the main temporal lobe of amplitude about its maximum, and ignores the temporal parts of the pulse of low amplitude. First, we consider the instantaneous frequency at the time $\tau=0$ of maximum amplitude, defined as
\begin{equation}
\bar\omega_c (\rho)= \left.\frac{d\,{\rm arg}\, E(\rho,\tau)}{d\tau}\right|_{\tau=0}\,,
\end{equation}
which will be called the central frequency to distinguish it from $\bar\omega(\rho)$, since their values are generally different. Writing ${\rm arg}\, E =\tan^{-1}({\rm Im}\,E/{\rm Re}\, E)$ with ${\rm Re}\, E =(E+E^\star)/2$ and ${\rm Im}\,E=(E-E^\star)/2i$, and using Eq. (\ref{AN2}) and its derivative with respect to time, one arrives at the alternate expression
\begin{equation}\label{MEANA}
\bar\omega_c(\rho) = \frac{\int_0^\infty \hat E_\omega(\rho) \omega d\omega}{\int_0^\infty \hat E_\omega(\rho) d\omega}=
\frac{\int_0^\infty |\hat E_\omega(\rho)| \omega d\omega}{\int_0^\infty |\hat E_\omega(\rho)| d\omega}\, .
\end{equation}
The introduction of the absolute values in the last step does not alter the result with the spectral phases in Eq. (\ref{LG3}) independent of frequency. In this way $\bar \omega_c$ is easily measurable, e. g., from the square root of the spectral density $|\hat E_\omega|^2$ provided by a spectrometer. Second, a measure of the width of the central lobe of the pulse based on its concavity is provided by the expression
\begin{equation}\label{DURATIONA}
\Delta t_c^2(\rho) \equiv -2 \left(\left.\frac{d^2|E(\rho,\tau)|/d\tau^2}{|E(\rho,\tau)|}\right|_{\tau=0}\right)^{-1} \,,
\end{equation}
that yields the $1/e$-decay half-duration for a Gaussian-shaped amplitude. For other pulse shapes Eq. (\ref{DURATIONA}) provides the duration of a Gaussian pulse with the same concavity of the maximum at $\tau=0$, and hence will be called the central duration. Using the equivalent expression $\Delta t_c^2 =-4|E|^2/(d^2|E|^2/d\tau^2|_{\tau=0})$, writing $|E|^2=E E^\star$, Eq. (\ref{AN2}) and its derivatives with respect to time at $\tau=0$, one easily arrives at the alternate expression
\begin{equation}\label{DURATIONA2}
\Delta t_c^2(\rho)=4/\Delta\omega_c^2(\rho)
\end{equation}
where
\begin{eqnarray}\label{BANDWIDTHA}
\Delta\omega_c^2(\rho) &=& 2\frac{\int_0^\infty \hat E_\omega(\rho) [\omega-\bar\omega_c(\rho)]^2 d\omega}{\int_0^\infty \hat E_\omega(\rho) d\omega} \nonumber \\
&=&
2\frac{\int_0^\infty |\hat E_\omega(\rho)| [\omega-\bar\omega_c(\rho)]^2 d\omega}{\int_0^\infty |\hat E_\omega(\rho)| d\omega}
\end{eqnarray}
defines a spectral bandwidth based on the spectral amplitude instead of the spectral intensity, yields, again, the $1/e$-decay half-bandwidth for a Gaussian-like spectral amplitude $|\hat E_\omega(\rho)|$, and therefore coincides with $\Delta \omega(\rho)$ for a Gaussian spectrum. The measures $\Delta \omega$ and $\Delta\omega_c$ are however different for other spectral shapes; in particular $\Delta \omega_c$ overweights widespread frequencies far from the central frequency of the spectrum compared to $\Delta\omega$. The two parameters $\omega_c$ and $\Delta t_c$ allow to reconstruct the shape of the central portion the pulse very reliably. Figure \ref{Fig2} shows two ultrashort pulse shapes and their approximation $e^{-\tau^2/\Delta t_c^2}e^{-i(\bar\omega_c\tau+\Phi)}$ to its central part of high amplitude based on the central frequency and duration.

\subsection{Lower bound to the ultrafast vortex duration at its most intense ring}

With the above pulse properties in mind, let us locate the caustic surface $\rho_E$ of the ultrafast vortex where the peak pulse amplitude is maximum. After some calculations, the derivative of the peak amplitude in Eq. (\ref{PEAK}) with respect to $\rho$ is found to be
\begin{equation}\label{DPEAK}
\frac{d|E(\rho,0)|}{d\rho} =  \frac{|E(\rho,0)|}{\rho} \left[|l| - 2\rho^2 \bar \omega_c(\rho)\right]\,.
\end{equation}
The fluence will then be maximum or minimum at the caustic surface determined by
\begin{equation}
\rho_E^2 = \frac{|l|}{2\bar\omega_c(\rho_E)}\,,
\end{equation}
which generally differs from the caustic surface $\rho_F^2 = |l|/2\bar\omega(\rho_F)$ of maximum fluence \cite{PORRAS_PRL}. Evaluation of the second derivative of $|E(\rho,0)|$ with respect to $\rho$ leads to an involved expression, which however at $\rho_E$ simplifies to
\begin{equation}\label{DDPEAK}
\left.\frac{d^2|E(\rho,0)|}{d\rho^2}\right|_{\rho_E} = -2\bar\omega_E(\rho_E)|E(\rho_E,0)| \left[2 - \frac{|l|}{2}\frac{\Delta\omega_c^2(\rho_E)}{\bar\omega_c^2(\rho_E)}\right]\,.
\end{equation}
The condition of maximum in Eq. (\ref{DDPEAK}) allows us to conclude that the bandwidth $\Delta\omega_c$ at the caustic surface of maximum intensity and the topological charge of ultrafast vortices always satisfy inequality
\begin{equation}\label{UPPERBOUND}
\frac{\Delta\omega_c(\rho_E)}{\bar\omega_c(\rho_E)} < \frac{2}{\sqrt{|l|}}\,,
\end{equation}
i. e., the relative bandwidth at the most intense ring cannot exceed a maximum value determined by the topological charge of the vortex. Inequality (\ref{UPPERBOUND}) is formally equal as $\Delta\omega(\rho_F)/\bar\omega(\rho_F)< 2/\sqrt{|l|}$, but these two inequalities refer to different caustic surfaces and the quantities involved are different. The above inequality implies a lower bound to the duration of the central lobe of an ultrafast vortex at its most intense ring. From Eq. (\ref{DURATIONA2}), inequality then (\ref{UPPERBOUND}) transforms into
\begin{equation}\label{LOWERBOUND}
\Delta t_c(\rho_E) > \frac{\sqrt{|l|}}{\bar\omega_c(\rho_E)}\,.
\end{equation}
This inequality imposes a lower bound to the central duration relative to the central period $2\pi/\bar\omega_c(\rho_E)$, and therefore a lower bound to the number of oscillations of the pulse at the most intense ring of the ultrafast vortex. If the number of oscillations $N$ is measured as the full width at half maximum of intensity of the central lobe, $\sqrt{2\ln 2} \Delta t_c(\rho_E)$, over the central period, inequality (\ref{LOWERBOUND}) leads to inequality $N>\sqrt{\ln 2 /2}\sqrt{|l|}/\pi$ for the number of oscillations. Since the lower bound in the right hand side increases monotonically with $|l|$, the number of oscillations in the central lobe of the pulse at the most intense ring of the ultrafast vortices synthesized with the same source spectrum $\hat a_\omega$ increases if the imprinted topological charge is increased. In practice, as seen below, the central frequency $\bar\omega_c(\rho_E)$ of the pulse at the most intense ring is substantially determined by the available frequencies in the superposition, i. e., by $\hat a_\omega$, indeed $\bar\omega_c(\rho_E)\approx \bar\omega_c$, with
\begin{equation}
\bar\omega_c = \frac{\int_0^\infty |\hat a_\omega| \omega d\omega}{\int_0^\infty |\hat a_\omega| d\omega}\,,
\end{equation}
With $\bar\omega_c(\rho_E)\approx \bar\omega_c$ independent of the topological charge, inequality (\ref{LOWERBOUND}) imposes directly a lower bound $\sqrt{|l|}/\bar\omega_c$ to the duration of the main lobe, and implies an increase of this duration with the magnitude of the topological charge.

\section{Ultrafast vortices with prescribed pulse shape at the most intense ring}

As a verification, suppose we need the expression an ultrafast vortex of certain transform-limited temporal shape,
\begin{equation}
P(\tau)= \frac{1}{\pi} \int_0^\infty \hat P_\omega e^{-i\omega\tau} d\omega \,,
\end{equation}
at the most intense ring, that is characterized by certain central frequency $\bar\omega_c$ and duration $\Delta t_c$. Equating (\ref{AN2}) with (\ref{LG2}) evaluated at $\rho_E^2=|l|/2\bar\omega_c(\rho_E) = |l|/2\bar \omega_c$ to $D P(\tau)$, the needed laser source spectrum must be $\hat a_\omega = \hat P_\omega\omega^{-|l|/2} e^{\rho_E^2\omega}(\sqrt{2}\rho_E)^{-|l|}$. Introducing this expression of $\hat a_\omega$ in Eq. (\ref{LG2}) for generic $\rho$, the result in Eq. (\ref{AN2}), and performing the integral, yields the expression
\begin{equation}\label{FIXED1}
E(\rho,\tau)= D \left(\frac{\rho}{\rho_E}\right)^{|l|} P\left(\tau-i\rho^2 + i\rho_E^2\right)\,,
\end{equation}
of the ultrafast vortex of pulse shape $P(\tau)$ at $\rho_E$. Coming back to the physical radius $r$, the caustic of maximum peak intensity is $r_E(z)=\sqrt{|l|/2}\, s_{\bar\omega_c}(z)$, where $s_{\bar\omega_c}(z)$ is given by Eqs. (\ref{S}) evaluated at $\bar\omega_c$, and the ultrafast vortex reads
\begin{equation}\label{FIXED2}
E = D\! \left(\sqrt{\frac{2}{|l|}}\frac{r}{s_{\bar\omega_c}(z)}\right)^{|l|}\!\!\!\!P\left[t'-\frac{r^2}{2cq(z)} + i\frac{|l|}{2\bar\omega_c}\right]\,.
\end{equation}
With given $P(\tau)$, and hence given $\bar\omega_c$ and $\Delta t_c$, Eq. (\ref{FIXED2}) represents indeed an ultrafast vortex with pulse shape $P(\tau)$ at is most intense ring at $r_E(z)=\sqrt{|l|/2}\, s_{\bar\omega_c}(z)$ only if $|l|< \bar\omega_c^2 \Delta t_c^2$; otherwise the intensity takes a relative minimum at $r_E(z)=\sqrt{|l|/2}\, s_{\bar\omega_c}(z)$, with a maximum at another radius with different pulse shape such that inequality (\ref{LOWERBOUND}) is satisfied, or possibly a singular expression, as in the following example.

\begin{figure}[b!]
\includegraphics*[width=6.5cm]{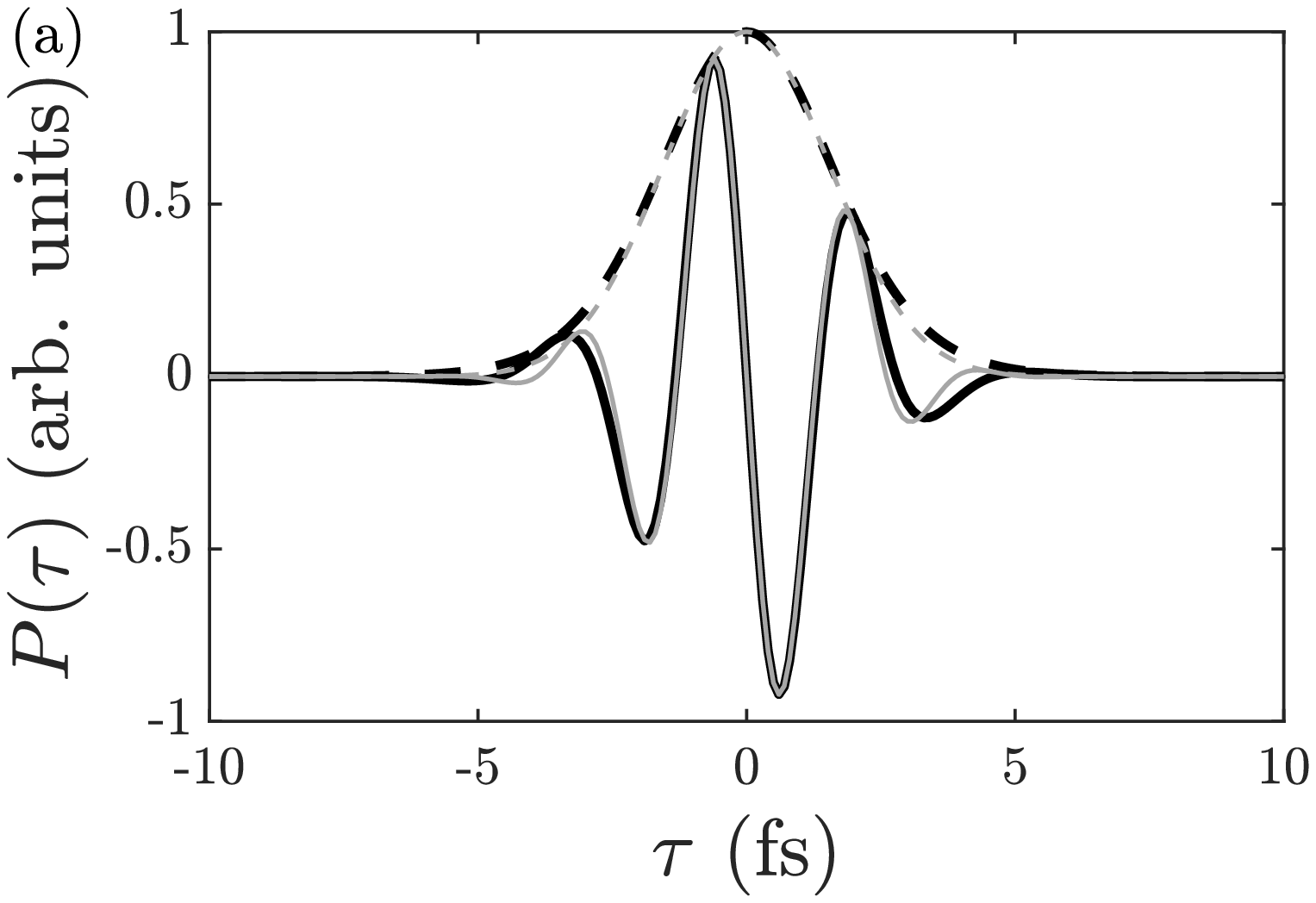}
\includegraphics*[width=6.5cm]{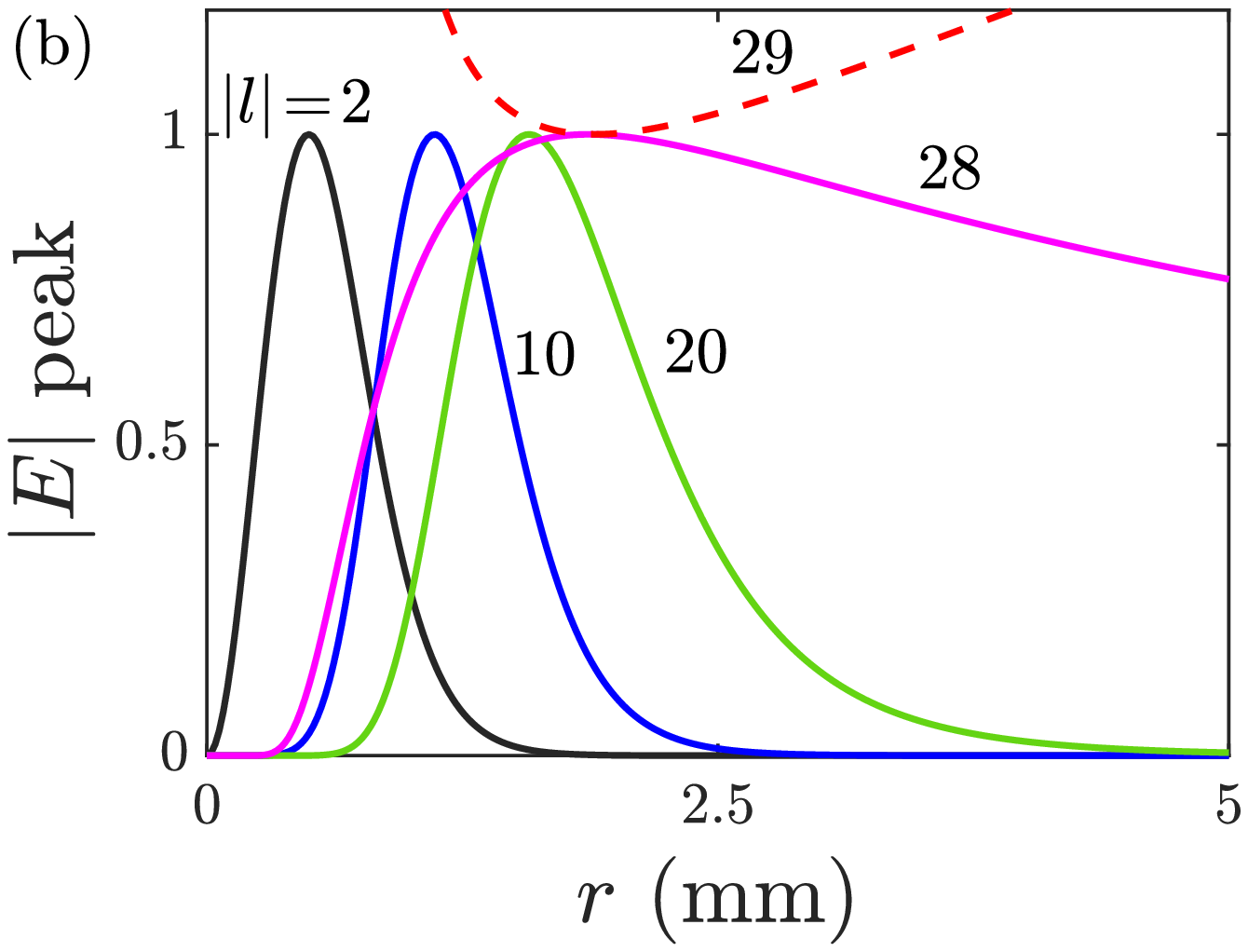}
\caption{(a) The real part (black solid) and amplitude (black dashed) of the pulse $P$ in Eq. (\ref{P}) with $\bar\omega_c=2.417$ rad/fs (780 nm wave length), $\beta=14.24$, and $\Phi=\pi/2$, containing a single oscillation within its full width at half maximum in intensity, and its approximation $P_c = e^{-\tau^2/\Delta t_c^2}e^{-i(\bar\omega_c\tau +\Phi)}$, with $\Delta t_c=2.208$ fs (gray curves). (b) Peak amplitude of the ultrafast vortices of Rayleigh distance $z_R=1$ m with the pulse shape $P(\tau)$ at $r_E(z)=\sqrt{|l|/2}\, s_{\bar\omega_c}(z)$ as a function of radius as given by Eq. (\ref{FIXEDPP}) at $z=0$ for increasing values of $l$, featuring maxima at $r_E(0)=\sqrt{|l|/2}\, s_{\bar\omega_c}(0)$ for $|l|<2\beta$ and minima at $r_E(0)=\sqrt{|l|/2}\, s_{\bar\omega_c}(0)$ for $|l|>2\beta$, with a singularity at a smaller radius, and thus lacking physical meaning.}
\label{Fig3}
\end{figure}

We consider the transform-limited pulses
\begin{equation}\label{P}
P(\tau)= \left(\frac{-i\beta}{\bar\omega_c\tau- i\beta}\right)^\beta e^{-i\Phi}\,,
\end{equation}
where $\bar\omega_c$ is just the central frequency, and the parameter $\beta\ge 1$ determines the central bandwidth $\Delta\omega_c=\sqrt{2/\beta}\,\bar\omega_c$ and the duration $\Delta t_c= \sqrt{2\beta}/\bar\omega_c$. For $\beta\gg 1$, $P(\tau)$ approaches the Gaussian pulse $\exp(-\tau^2/\Delta t_c^2)e^{-i(\bar\omega_c\tau+\Phi)}$ based on its central parameters not only in its central portion but at any time because $P(\tau)$ approaches a Gaussian-enveloped pulse, as in the example of Fig. \ref{Fig3}(a) showing a near infrared, single-cycle pulse ($\bar\omega_c=2.417$ rad/fs, $\beta=14.24$). Condition $|l|<\bar\omega_c^2 \Delta t_c^2$ for the existence of the ultrafast vortex with this pulse shape at its most intense ring reads $|l|< 2\beta$. Indeed Eq. (\ref{FIXED2}) with Eq. (\ref{P}) yields
\begin{eqnarray}\label{FIXEDP}
E &=& D \left(\sqrt{\frac{2}{|l|}}\frac{r}{s_{\bar\omega_c}(z)}\right)^{|l|} \nonumber \\
&\times& \left[\frac{-i\beta}{\bar\omega_c\left(t'\!-\!\frac{r^2}{2cq(z)}\right)\!+\! i\left(\frac{|l|}{2}\!-\!\beta\right)}\right]^\beta e^{-i\Phi} .
\end{eqnarray}
Its peak amplitude at $\tau=0$ [$t'=r^2/2cR(z)$],
\begin{eqnarray}\label{FIXEDPP}
\left|E(\tau=0)\right| &=& D\left(\sqrt{\frac{2}{|l|}}\frac{r}{s_{\bar\omega_c}(z)}\right)^{|l|} \nonumber \\
&\times & \left(\frac{\beta}{\frac{r^2}{s_{\bar\omega_c}^2(z)}-\frac{|l|}{2} + \beta}\right)^\beta
\end{eqnarray}
as a function of $r$ is depicted in Fig. \ref{Fig3}(b) for increasing values of $|l|$, and is characterized by a maximum at the corresponding radii $r_E(z)=\sqrt{|l|/2}\, s_{\bar\omega_c}(z)$ if $|l|<\bar\omega_c^2 \Delta t_c^2$ (i. e., if $|l|<2\beta=28.48$), but by a minimum at $r_E(z)=\sqrt{|l|/2}\, s_{\bar\omega_c}(z)$ if $|l|> \bar\omega_c^2 \Delta t_c^2$ (i. e., $|l|>2\beta=28.48$) because the peak amplitude in Eq. (\ref{FIXEDPP}) has a singularity at a radius smaller than $r_E(z)$, which makes Eq. (\ref{FIXEDP}) physically invalid. Thus, an ultrafast vortex whose pulse shape at its most intense ring is a single-cycle pulse can carry only up to $28$ units of OAM.

\subsection{Ultrafast vortices with power-exponential Laguerre-Gauss spectrum}

In practice one has a spectrum $\hat a_\omega$ that is substantially determined by the laser source and is therefore independent of the topological charge $l$, and with this spectrum may wish to synthesize an ultrafast fortex of certain charge $l$. The above OAM-temporal coupling effects imply that the temporal properties of the main lobe of the pulse at the most intense ring of the ultrafast vortices synthesized with the same spectrum $\hat a_\omega$ change with $l$. This is particularly true if the bandwidth of $\hat a_\omega$ is very large and/or the topological charge high.

\begin{figure*}[tbp]
\includegraphics*[width=5.9cm]{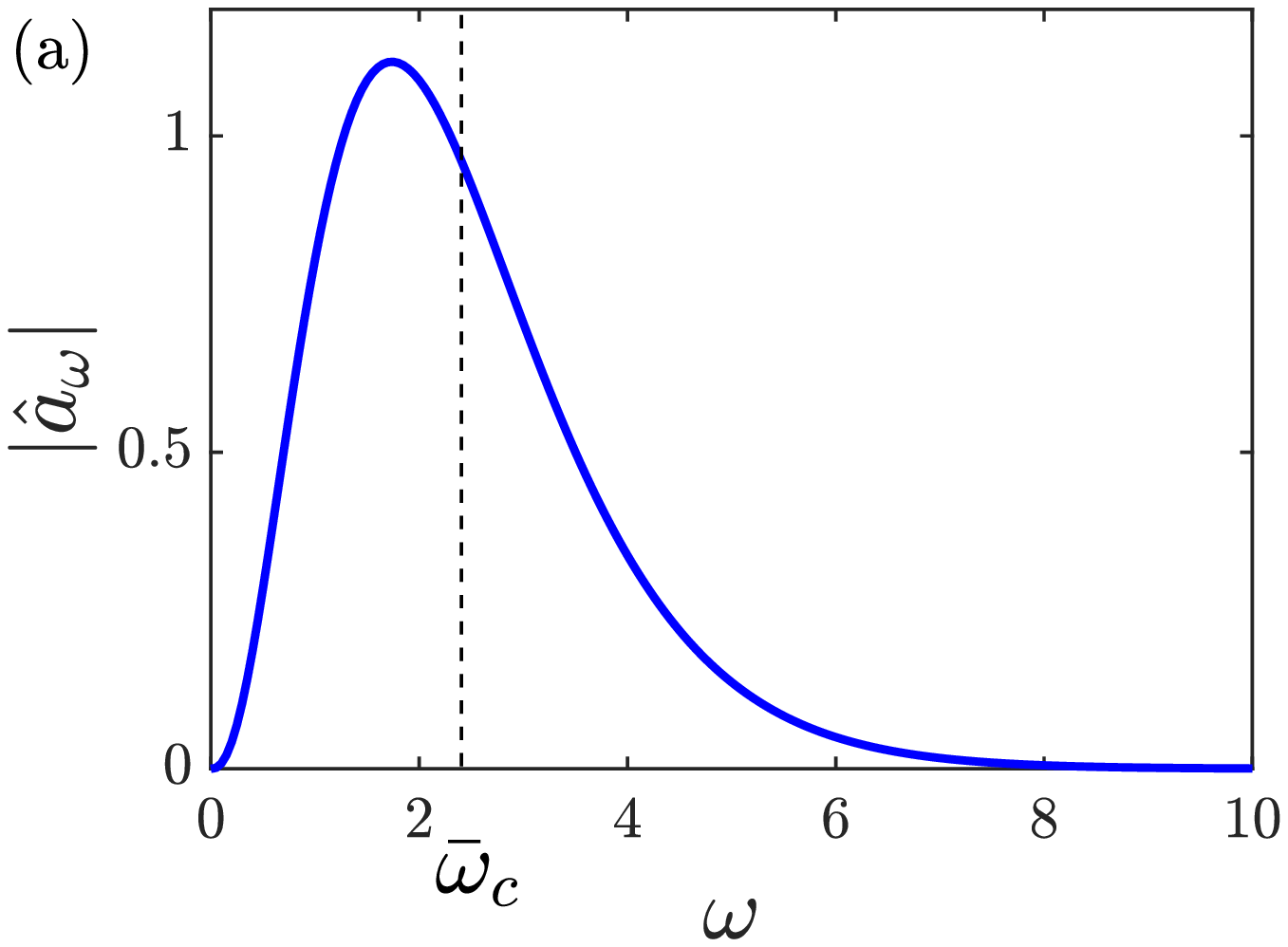}
\includegraphics*[width=5.9cm]{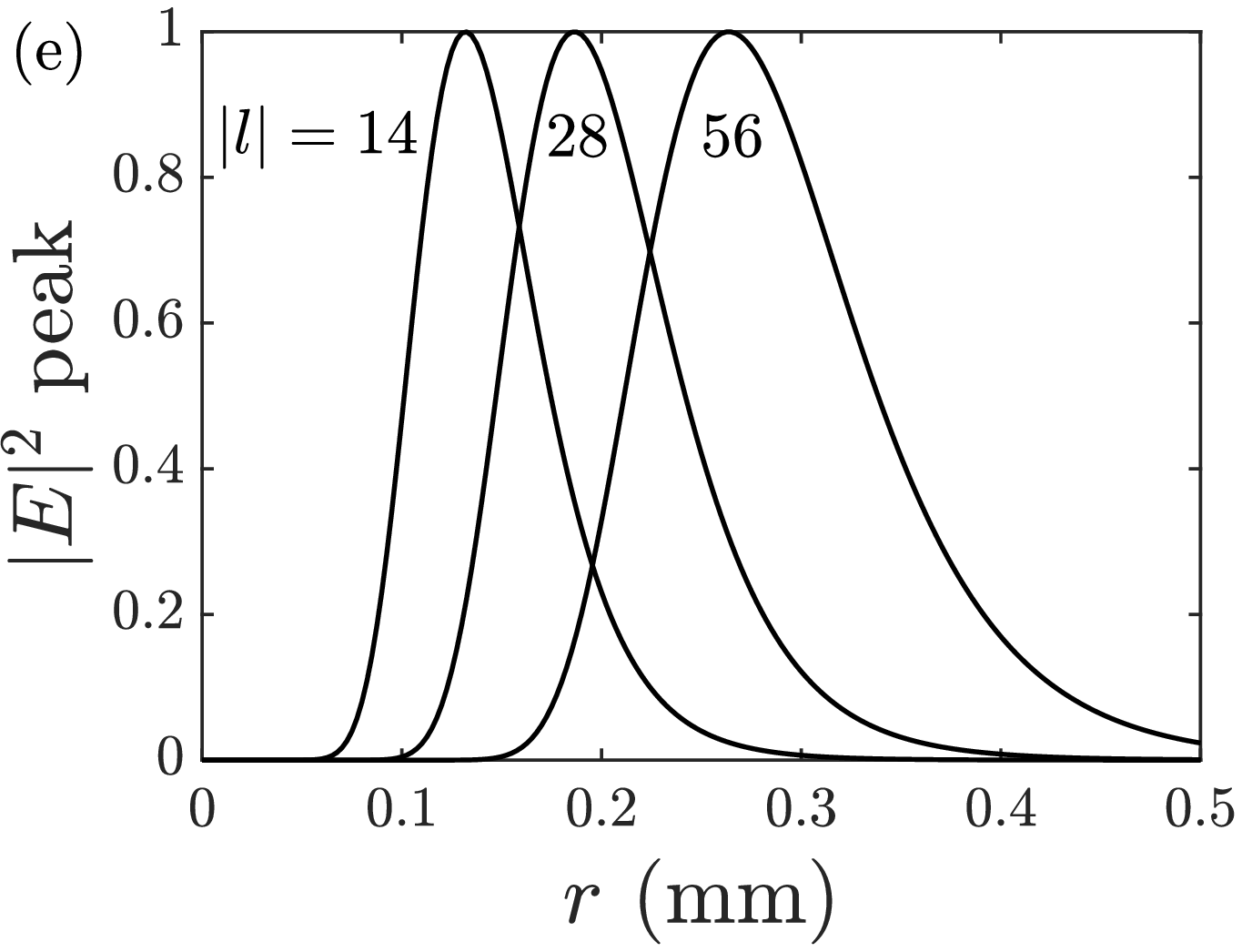}
\includegraphics*[width=5.9cm]{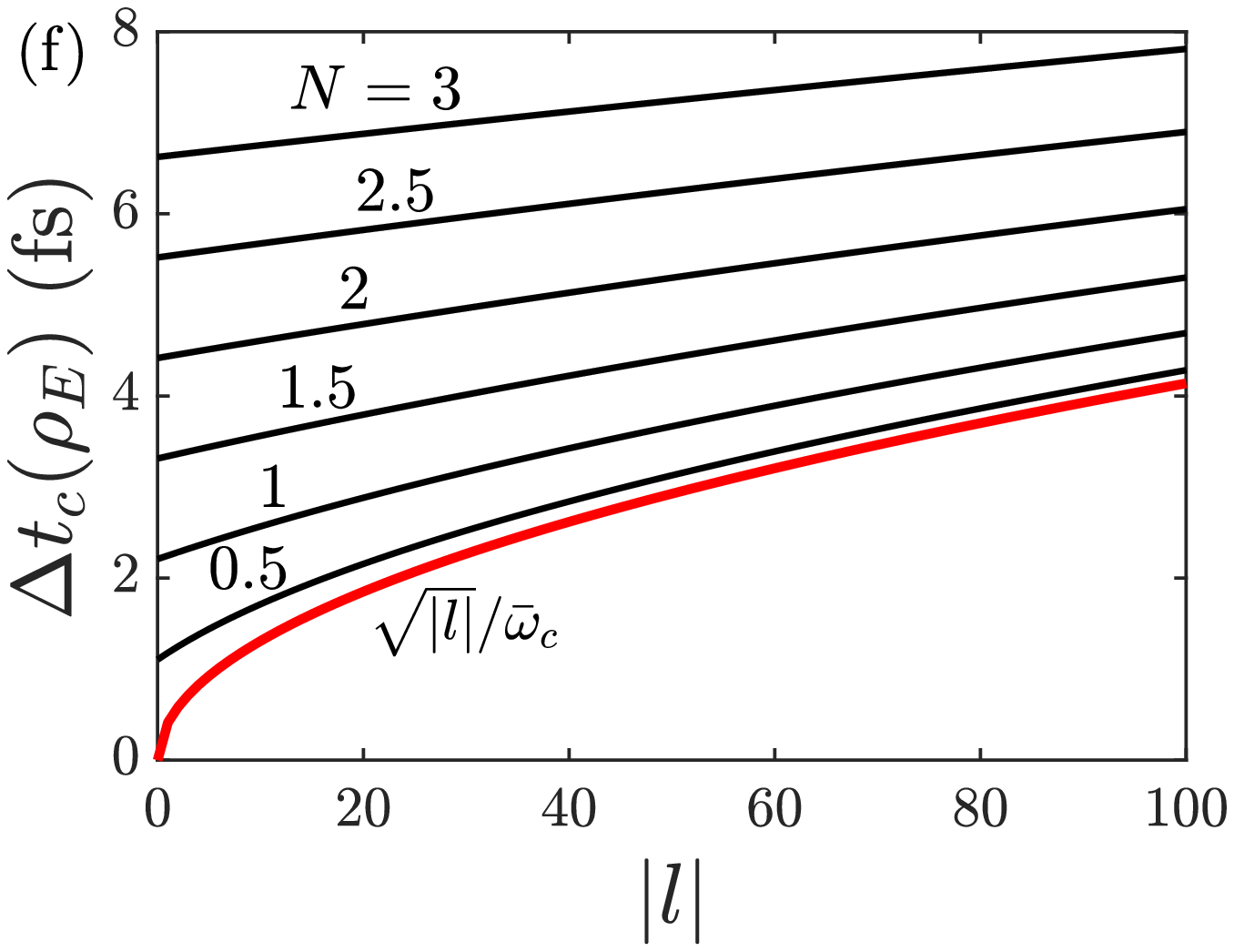}
\includegraphics*[width=5.9cm]{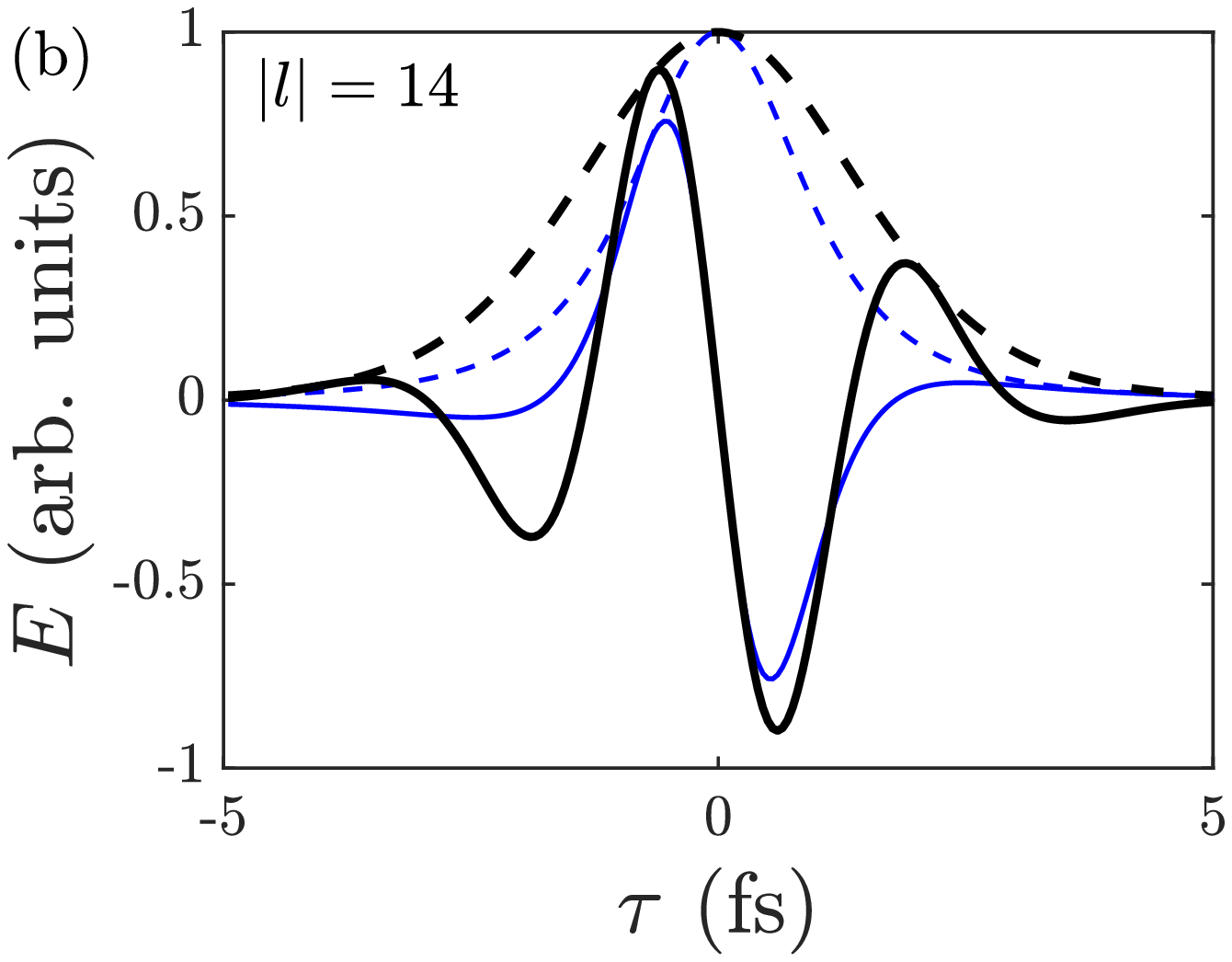}
\includegraphics*[width=5.9cm]{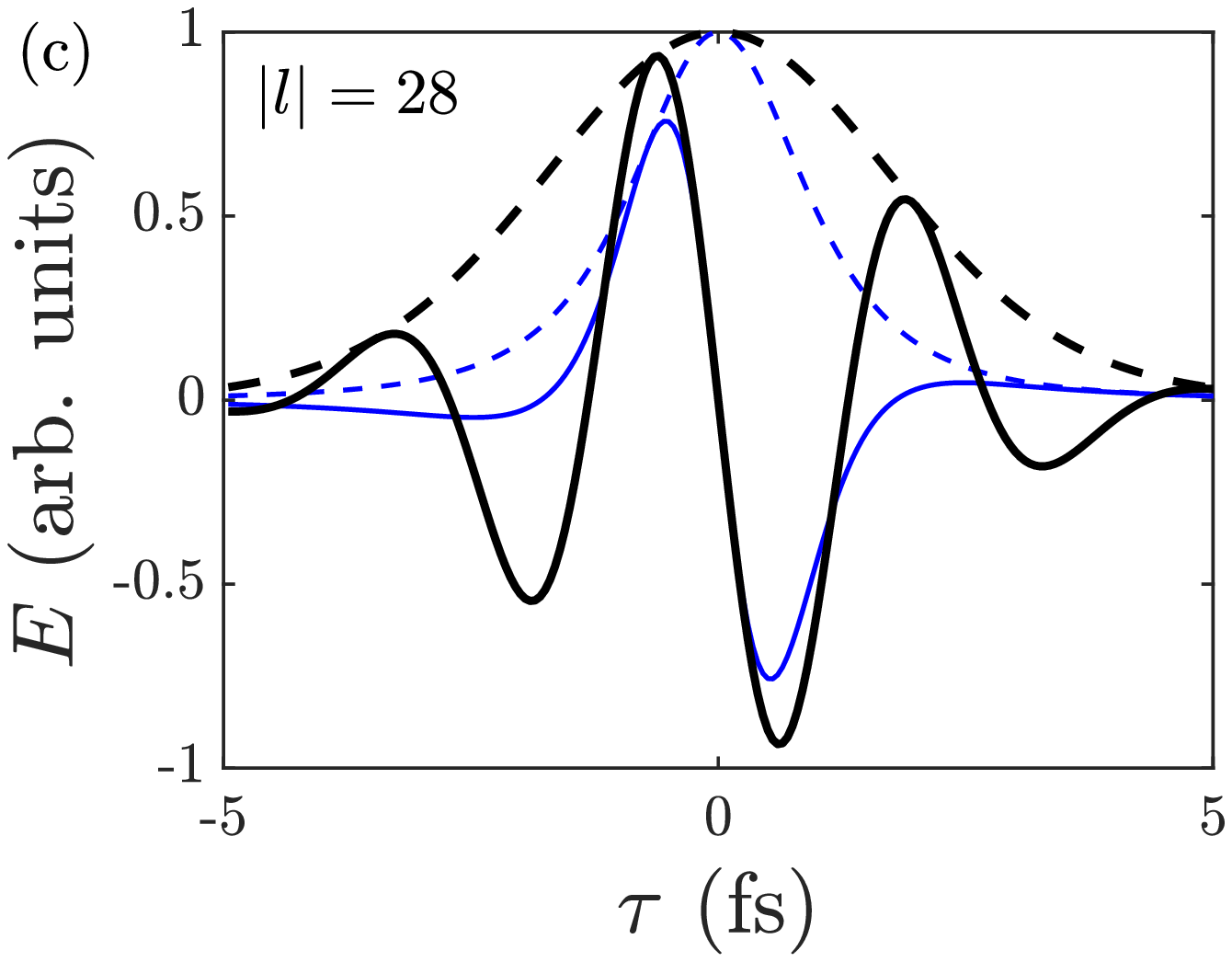}
\includegraphics*[width=5.9cm]{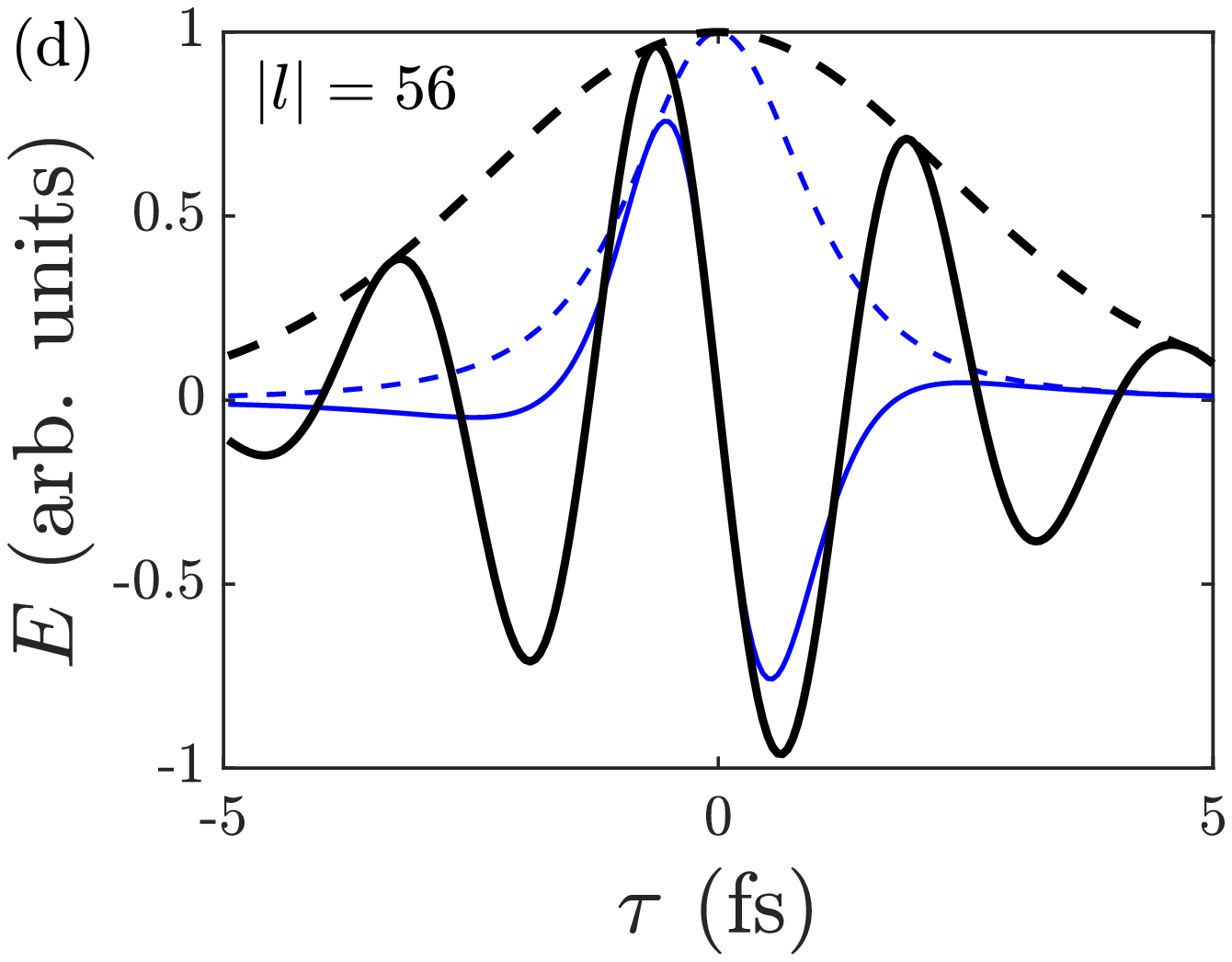}
\caption{(a) Power-exponential spectrum $\hat a_\omega$ in Eq. (\ref{PE}) with $\bar\omega_c=2.417$ rad/fs and $\beta=3.56$. (b,c,d) Real part and amplitude of the half-cycle ($N=0.5$) pulse $a(\tau)$ in time domain (blue curves), and real part and amplitude of the pulse at the most intense ring of ultrafast vortices with the indicated values of $|l|$, as given by Eq. (\ref{NOFIXEDPULSE}). All pulse shapes are normalized to their peak values for a better comparison. (e) With $z_R=10$ mm, peak intensity of the pulse as a function of radius at the waist $z=0$ for the three values of $|l|$ with maxima at $r_E(0)=\sqrt{|l|/2}\, s_{\bar \omega_c}$. (f) Central duration of vortices at their most intense ring as a function of the magnitude of the topological charge $|l|$, for power-exponential laser spectra with $\bar\omega_c=2.417$ rad/fs, and with $\beta=3.56$ ($N=0.5$), $\beta= 14.24$ ($N=1$), $\beta=32.04$ ($N=1.5$), $\beta=56.95$ ($N=2$), $\beta=89.0$ ($N=2.5$) and $\beta=128.15$ ($N=3$). The red curve is the lower bound to the pulse duration in inequality (\ref{LOWERBOUND}).}
\label{Fig4}
\end{figure*}

As a sufficiently flexible model, we take the power-exponential spectrum
\begin{equation}\label{PE}
\hat a_\omega= \frac{\pi}{\Gamma(\beta)}\left(\frac{\beta}{\bar\omega_c}\right)^{\beta-1} \bar\omega_c^\beta
e^{-\frac{\omega}{\bar\omega_c}\beta}e^{-i\Phi}\,,
\end{equation}
corresponding in time domain to the same pulse $a(\tau)=[-i\beta/(\bar\omega_c \tau -i\beta)]^\beta e^{-i\Phi}$ as $P(\tau)$ in Eq. (\ref{P}) of arbitrary central frequency $\bar\omega_c$ and central duration $\Delta t_c =\sqrt{2\beta}/\bar\omega_c$. Integral in Eq. (\ref{AN}) with the LG beams in Eq. (\ref{AN}) and the power-exponential spectrum in Eq. (\ref{PE}) can be carried out to yield the closed-form expression
\begin{eqnarray}\label{NOFIXED}
E &=& D \left(\sqrt{\frac{2}{|l|}}\frac{r}{s_{\bar\omega_c}(z)}\right)^{|l|} \nonumber \\
&\times& \left[\frac{-i\left(\beta+\frac{|l|}{2}\right)}{\bar\omega_c\left(t'-\frac{r^2}{2cq(z)}\right) - i\beta}\right]^{\beta+\frac{|l|}{2}} e^{-i\Phi}
\end{eqnarray}
which is, this time, nonsingular and localized for any value of $l$. A multiplicative factor has been introduced in Eq. (\ref{NOFIXED}) to adjust to unity the peak amplitude at the ring $r_E(0)=\sqrt{|l|/2}\, s_{\bar\omega_c}$ at the waist $z=0$. The peak amplitude at any point $(r,z)$ is given by
\begin{eqnarray}\label{NOFIXEDPP}
\left|E(\tau=0)\right| &=& D\left(\sqrt{\frac{2}{|l|}}\frac{r}{s_{\bar\omega_c}(z)}\right)^{|l|} \nonumber \\
&\times & \left(\frac{\beta+\frac{|l|}{2}}{\beta + \frac{r^2}{s_{\bar\omega_c}^2(z)}}\right)^{\beta+\frac{|l|}{2}}
\end{eqnarray}
which takes a maximum value at the caustic surface $r_E(z) =\sqrt{|l|/2}\, s_{\bar\omega_c}(z)$. At this caustic surface the pulse shape is
\begin{equation}\label{NOFIXEDPULSE}
E(\tau) = D\left[\frac{-i\left(\beta+\frac{|l|}{2}\right)}{\bar\omega_c \tau - i\left(\beta+\frac{|l|}{2}\right)}\right]^{\beta+\frac{|l|}{2}} e^{-i\Phi}\,,
\end{equation}
which is the same as $a(\tau)$ but with $\beta$ replaced with $\beta+|l|/2$. Thus, for the fundamental Gaussian beam ($l=0$), i. e., the fundamental isodiffracting pulsed Gaussian beam \cite{PORRAS_PRE,PORRAS_JOSAB}, the pulse shape at $r=0$ is just $a(\tau)$ of central frequency $\bar\omega_c$ and duration $\Delta t_c=\sqrt{2\beta}/\bar\omega_c$, as determined by the full source spectrum $\hat a_\omega$. For $|l|\neq 0$, the pulse temporal shape at the most intense ring maintains the frequency $\bar\omega_c(r_E)=\bar\omega_c$ of the source spectrum regardless of the value of $l$, but the duration increases with the magnitude of the topological charge as
\begin{equation}
\Delta t_c(r_E)= \frac{\sqrt{2\beta + |l|}}{\bar\omega_c} \,;
\end{equation}
in particular, pulse adapts itself so that its duration is always above the lower bound $\sqrt{|l|}/\bar\omega_c(r_E) = \sqrt{|l|}/\bar\omega_c$. Figure \ref{Fig4} illustrates the adaptation of pulse shape at the most intense caustic surface to the value of the topological charge. With the ultra-broadband spectrum of Fig. \ref{Fig4}(a) of central frequency in the near infrared, the half-cycle pulse represented by blue curves in Figs. \ref{Fig4}(b), (c) and (d) could be built with $l=0$ at $r=0$. At the maxima of the radial distribution of peak intensity depicted in Fig. \ref{Fig4}(e) for several values of $|l|$, the pulse shapes represented in Figs. \ref{Fig4}(b), (c) and (d) as black curves are seen to increase their duration with increasing $|l|$. As a summary of this OAM-temporal coupling effect, the black curves in Fig. \ref{Fig4}(f) represent the pulse duration as a function of the topological charge for increasing bandwidths of $\hat a_\omega$ that correspond to laser pulses with increasing number $N$ of cycles. Irrespective of the source bandwidth and associated number of cycles, the duration at the most intense ring of the produced ultrafast vortices is always above $\sqrt{|l|}/\bar\omega_c$, the adaptation being more pronounced as the number of cycles is smaller and vanishing in the monochromatic limit.

\section{Conclusions}

In conclusion, we have described how the OAM-temporal coupling in ultrafast vortices affects the temporal properties of the vortex at the ring of maximum peak intensity, which is the ring of primary relevance for many nonlinear optics applications, particularly those involving strong-field vortex-matter interactions. In a few words, the ultrafast vortices synthesized with a given laser source spectrum increase their duration if the imprinted topological charge is increased in order that the pulse duration at the most intense ring is always above a lower bound that increases monotonically with the magnitude of the topological charge.  We have provided analytical expressions of ultrafast vortices that incorporate these effects and can be used as a starting point for the study of their linear and nonlinear interaction with matter.

Although we have limited our considerations to ultrafast vortices in optics, the same OAM-temporal coupling effects affect ultrafast electron vortices \cite{MCMORRAN} and transient acoustic vortices \cite{MARZO}.

The author acknowledges support from Projects of the Spanish Ministerio de Econom\'{\i}a y Competitividad No. PGC2018-093854-B-I00, and No. FIS2017-87360-P.

\end{document}